\documentclass{iau}
\usepackage{graphicx}

\title[The XMM-Newton spectral-fit database] 
{The XMM-Newton spectral-fit database}

\author[A. Corral, I. Georgantopoulos, S. Rosen, M.G. Watson, K. Page, \& G.C. Stewart]   
{A. Corral$^1$, I. Georgantopoulos$^1$, S. Rosen$^2$, M.G. Watson$^2$, K. Page$^2$, \and G.C. Stewart$^2$}

\affiliation{$^1$Institute for Astronomy, Astrophysics, Space Applications, and Remote Sensing (IAASARS), National Observatory of Athens (NOA),\\ I.Metaxa \& Vas. Pavlou St., 15236, Penteli, Greece \\ email: {\tt acorral@noa.gr}\\ [\affilskip] $^2$X-ray and Observational Astronomy Group (XROA), Department of Physics and Astronomy, University of Leicester,\\ University Road, Leicester, LE1 7RH, United Kingdom} 

\pubyear{2013}
\volume{304}  
\pagerange{119--126}
\jname{Multiwavelength AGN Surveys and Studies}
\editors{A. Mickaelian, F. Aharonian \& D. Sanders.}
\begin{document}

\maketitle

\begin{abstract}
The XMM-Newton spectral-fit database is an ongoing ESA funded project
aimed to construct a catalogue of spectral-fitting results for all the
sources within the XMM-Newton serendipitous source catalogue for which
spectral data products have been pipeline-extracted ($\gtrsim$ 120,000
X-ray source detections). The fundamental goal of this project is to
provide the astronomical community with a tool to construct large and
representative samples of X-ray sources by allowing source selection
according to spectral properties.

\keywords{X-ray surveys.}
\end{abstract}

\firstsection 
\section{Introduction}

The XMM-Newton serendipitous source catalogue is the largest catalogue
of X-ray sources built to date, see \cite[Watson
  \etal\ (2009)]{watson}. In its latest version, the 3XMM-Newton Data
Release 4 (3XMM
DR4\footnote{http://xmmssc-www.star.le.ac.uk/Catalogue/3XMM-DR4/}), it
contains photometric information for more than 500,000 source
detections, corresponding to $\sim$ 370,000 unique sources. Besides,
spectra and time series were also extracted if the number of source
counts collected by the EPIC camera was $>$ 100 counts. The resulting
catalogue contains spectra for more than 120,000 detections
corresponding to $\sim$ 85,000 unique sources. The project described
here, is aimed to take advantage of the great wealth of data and
information contained within the XMM-Newton source catalogue, to
construct a database composed of spectral-fitting results.

\section{Automated spectral-fitting procedure}

The XMM-Newton spectral-fit database is constructed by using automated
spectral fits applied to the pipe-line extracted spectra within the
3XMM DR4 catalogue. The software used to perform the spectral fits is
{\tt XSPEC v12.7} (see \cite[Arnaud (1996)]{arnaud}). The statistics
used to fit the data is Cash statistics, implemented as C-stat in {\tt
  XSPEC}. This statistics was selected, to optimise the spectral
fitting in the case of low quality spectra. However, even using this
statistics, a lower limit on the number of counts in each individual
spectrum of 50 counts per instrument had to be imposed to ensure a
minimum quality on the spectral fits. Therefore, {\bf the spectral-fit
  database is composed of spectral-fitting results for $\gtrsim$
  114,000 detections, corresponding to $\sim$ 77,000 unique
  sources}.\\

Three energy bands are considered during the spectral fits: Full/Total
band (0.5 - 10 keV), Soft band (0.5 - 2 keV), and Hard band (2 - 10
keV). Six spectral models have been implemented in the
spectral-fitting pipe-line: three simple models (models 1 to 3), and
three more complex models (models 4 to 6). These six models were
designed and optimised to reproduce the most commonly observed X-ray
spectral shapes among different astronomical sources. The three simple
models are applied to all sources provided that the number of counts
collected in the energy band in use is $>$ 50 counts:

{\bf 1. Absorbed power-law model} ({\tt wa*pow}): A
photoelectrically absorbed power-law model. This model is applied in
the three energy bands.

{\bf 2. Absorbed thermal model} ({\tt wa*mekal}): A
photoelectrically absorbed thermal model. This model is applied in the
total and soft bands.

{\bf 3. Absorbed black-body model} ({\tt wa*bb}): A photoelectrically
absorbed black-body model. This model is only applied in the soft
band, and its primary use is to obtain initial input parameters for a
more complex model.

The three more complex models are only applied if the number of counts
collected in the total band is $>$ 500 counts ($\gtrsim$ 28,000 detections,
corresponding to $\sim$ 18,000 unique sources):

{\bf 4. Absorbed power-law model plus thermal model} ({\tt
  wa(mekal+wa*pow)}): An absorbed thermal model plus a more highly
absorbed power-law model.  

{\bf 5. Double power-law model} ({\tt wa*(pow+wa*pow)}):. Two
power-law models with different photon indices, and additional
absorption for the second power-law component.

{\bf 6. Black-body plus power-law model} ({\tt wa*(bb+pow)}):. A
  black-body plus a power-law component, both absorbed by the same
  amount of gas.

\section{Automated spectral-fitting results}

A full description of this project as well as the full spectral-fit
database is presented in the webpage of this project:\\

{\bf http://xraygroup.astro.noa.gr/Webpage-prodec/index.html}. \\

The spectral-fit database contains one row per source and observation,
listing source information, and spectral-fit output parameters and
errors, as well as fluxes and additional information about the
goodness of fit for every model applied. In this way, users can
construct large and representative samples of X-ray sources by
querying this database according to spectral properties. In addition,
and in order to test the reliability of the automated fits, the
spectral-fitting process has also been applied to a sample of X-ray
sources extracted from the SDSS/XMM-Newton cross-correlation presented
in \cite[Georgakakis \& Nandra (2011)]{georg}. The spectral-fitting
pipe-line was modified to include the effect of redshift and Galactic
absorption, in the case of sources with either spectroscopic or
photometric redshifts within that sample.\\

 
\noindent
{\underline {\it Acknowledgements:}} A. Corral acknowledges financial
support by the European Space Agency (ESA) under the PRODEX program.

\end{document}